\documentclass{article}
\usepackage{graphicx} 

\usepackage{enumitem}
\usepackage{natbib}
\usepackage{booktabs}
\usepackage{array}
\usepackage{xcolor}
\usepackage{colortbl}
\usepackage{threeparttable}
\usepackage{amsthm}
\usepackage{fancyhdr}
\usepackage{microtype}
\usepackage{parskip}
\usepackage{amsmath}
\usepackage{adjustbox}
\usepackage{tabularx}
\usepackage{rotating}
\usepackage{babel}
\usepackage[margin=1in]{geometry}

\usepackage{pdflscape} 
\usepackage{appendix}

\title{Intraday Gas Fee Heterogeneity on Ethereum: Evidence from Operational Firms}
\author{
  Irene Aldridge\textsuperscript{1},
  Gavhar Annaeva\textsuperscript{1},
  Leyla Beriker\textsuperscript{1},
  Zhiheng Cai\textsuperscript{1},\\
  Samyak Choudhary\textsuperscript{1},
  Camila Godoy\textsuperscript{1},
  Kaicheng Gong\textsuperscript{1},
  Zitao Huang\textsuperscript{1},
  Jonah Ji\textsuperscript{1},\\
  Hetvi Kharvasiya\textsuperscript{1},
  Heng Li\textsuperscript{1},
  Yuxuan Li\textsuperscript{1},
  Tianchi Ma\textsuperscript{1},
  Qingcheng Meng\textsuperscript{1},\\
  Ruiyang Shi\textsuperscript{1},
  Ananya Shrivastava\textsuperscript{1},
  Jiaqi Wang\textsuperscript{1},
  Yifan Wang\textsuperscript{1},
  Zihua Wu\textsuperscript{1},\\
  Jiayang Xu\textsuperscript{1},
  Yuheng Yan\textsuperscript{1},
  Zijun Zeng\textsuperscript{1},
  Bowen Zhang\textsuperscript{1},
  Francesco Zhang\footnote{
  Johns Hopkins University -- Carey School of Business\\
  \texttt{irene.aldridge@gmail.com}
}
}

\date{}

\begin{document}

\maketitle

\begin{abstract}
Ethereum's EIP-1559 fee mechanism was designed under the assumption of
homogeneous, myopic agents responding to a single congestion signal.
We examine how this assumption interacts with the heterogeneous demand
structure of real-world Ethereum users. Analyzing 62,142 confirmed transactions from seven operational firms across seven industries (January--March 2026), we document significant intraday gas-fee variation: fees peak at hour~12 UTC (7\,AM ET, $\hat{\beta}_{12}=\$0.054$ above the U.S.\ evening baseline,
$p<0.001$) and are associated with periods of elevated
speculative-arbitrage activity. Operational firms exhibit heterogeneous scheduling responses moderated by transaction deferrability and gas intensity. Residual cost floors, i.e. the gap between observed expenditure and the counterfactual under perfect off-peak scheduling, range from 40.7\% to 92.5\% of actual expenditure, and persist even during the lowest-cost hours ($h\in\{20,21,22,23\}$ UTC, 3--6\,PM ET).
We introduce an On-Chain Scheduling Matrix that maps firms to four
scheduling regimes as a practical framework for managing gas-fee
exposure under the current mechanism.

\end{abstract}


\section{Introduction}
\label{sec:intro}
 
Blockchain networks impose execution costs through congestion pricing.
Ethereum's EIP-1559 mechanism \citep{buterin2021eip} raises the
protocol-determined base fee whenever aggregate block demand exceeds
a 50\% target utilization and lowers it otherwise.
\cite{roughgarden2021} establishes that this mechanism is
incentive-compatible for myopic, value-maximizing agents: the dominant
strategy is to bid true value, and the resulting allocation is
ex-post efficient under independent private values.
 
The theoretical guarantee applies under specific scope conditions,
most notably that agents are symmetric and respond myopically
to a common congestion signal. Ethereum's user population, however, includes at least two observationally distinct demand types.
\emph{Speculative-arbitrage actors} (MEV bots, DEX routers,
liquidation searchers) are highly price-elastic and cluster activity
during U.S.\ and European business hours, when arbitrage returns
are largest. \emph{Operational firms}, such as remittance platforms, healthcare
coordinators, real-estate settlement systems, face contractual,
counterparty, or regulatory timing constraints that limit their
ability to respond to real-time price signals. These two types share network infrastructure but interact with the base-fee mechanism in qualitatively different ways.
 
This paper documents the cost implications of that heterogeneity.
Using transaction-level on-chain data from seven firms across seven
industries ($N=62{,}142$ transactions, January--March 2026), we make
four contributions. First, we provide the first cross-industry empirical characterization of intraday gas-fee variation from the perspective of operational users. Second, we identify the boundary conditions under which the EIP-1559 incentive-compatibility result extends or does not extend to operationally constrained participants.
Third, we introduce a dual classification of gas fees as
\emph{execution costs in timing} (the firm chooses when to submit)
and \emph{maladaptation costs in origin} (the base fee is set by
aggregate network demand outside the firm's control), extending
Transaction Cost Economics \citep{williamson1985} to the blockchain
setting. Fourth, we develop an On-Chain Scheduling Matrix that translates the empirical fee schedule into actionable scheduling guidance across four firm-level regimes.
 
\section{Boundary Conditions of the EIP-1559 Fee Mechanism}
\label{sec:theory}
 
\subsection{The EIP-1559 Mechanism and Its Scope}
 
Under EIP-1559, the base fee for block $b$ adjusts deterministically
from the preceding block:
\begin{equation}
  B_b = B_{b-1} \cdot \left(1 + 0.125\cdot
        \frac{\phi_{b-1} - 0.5}{0.5}\right),
  \label{eq:basefee}
\end{equation}
where $\phi_b = G^{\text{used}}_b / G^{\text{max}}_b \in [0,1]$
is block fullness and the 12.5\% per-block adjustment cap limits
fee volatility. cite{roughgarden2021} proves truth-telling is a dominant strategy
for myopic agents with independent private values, yielding ex-post
efficient allocation.
The proof assumes agents evaluate each block independently and can
freely choose whether to transact based on their valuation versus
the prevailing base fee.
 
\subsection{Boundary Conditions Not Met in the Operational Setting}
 
Four assumptions underlying the theoretical result warrant empirical
examination in the context of operational firms.
 
\textbf{Assumption 1: Agent homogeneity.}
The model treats all agents as symmetric valuation-holders.
In practice, speculative-arbitrage actors are highly price-elastic
(entering when expected returns exceed fees) while operational firms
face near-inelastic demand within their scheduling windows.
This bimodal structure is not accommodated by the symmetric model.
 
\textbf{Assumption 2: Myopic, block-by-block optimization.}
The incentive-compatibility proof applies to agents choosing bids
one block at a time.
Operational firms typically schedule transactions days or weeks in
advance under contractual or regulatory constraints.
Their decision is not a fee bid but a timing commitment, placing
them outside the model's strategic framework.
 
\textbf{Assumption 3: Efficient clearing under the 50\% target.}
Because speculative demand is bursty and concentrated in business
hours globally, aggregate demand systematically exceeds the 50\%
target during those hours, raising base fees for all concurrent
users regardless of their individual demand characteristics.
Operational users who cannot defer transactions bear this congestion
externality without having contributed to it.
 
\textbf{Assumption 4: Externality resolution through pricing.}
The mechanism aims to clear congestion through price adjustment.
However, even during the lowest-congestion hours in our sample,
block-reward fullness $\hat{\phi}^{\text{br}}_{i,h^*}$ ranges from
0.184 to 0.462 across firms (Table~\ref{tab:floor}).
A non-zero residual congestion level persists throughout the day,
representing an unresolved cost burden for operational users that
scheduling discipline alone cannot eliminate.
 
\subsection{Extended TCE Classification}
 
Following \cite{williamson1985}, transaction costs decompose into
ex-ante, execution, and ex-post phases.
Standard TCE treats execution costs as stable conditional on
governance structure; this assumption does not hold on public
blockchains, where the total transaction fee
\begin{equation}
  C_{it} = g_{it}\cdot\bigl[B_t + \min(\pi_{it},\, M_{it}-B_t)\bigr]
  \label{eq:fee}
\end{equation}
contains a congestion component $g_{it}\cdot B_t$ determined by
aggregate network demand at time $t$, outside the firm's control.
We therefore propose a dual classification: gas fees are
\emph{execution costs in timing} (the firm can influence cost
through submission timing) and \emph{maladaptation costs in origin}
(the base fee reflects third-party demand the firm cannot
renegotiate or contractually insulate itself from).
This distinction extends the original TCE framework and is absent
from prior blockchain-in-operations studies
\citep{babich2020,lumineau2021}.
 
\section{Framework and Estimation}
\label{sec:methods}
 
\subsection{Intraday Cost Model}
 
The primary estimating equation regresses realized gas fees on
hour-of-day indicators in a firm-and-week fixed-effects panel:
\begin{equation}
  \text{Gas}_{it} = \alpha + \sum_{h=0}^{22} \beta_h\,
  \mathbf{1}[H_{it}=h]
  + \hat{\delta}\,\hat{\phi}^{\text{br}}_b
  + \mu_i + \lambda_w + \varepsilon_{it},
  \label{eq:main}
\end{equation}
where $\text{Gas}_{it}$ is the realized gas cost in USD;
$\mathbf{1}[H_{it}=h]$ is an indicator for UTC submission hour $h$,
with $h=23$ as the omitted baseline; $\hat{\phi}^{\text{br}}_b$
is a block-reward fullness proxy (defined below);
$\mu_i$ are firm fixed effects; $\lambda_w$ are week fixed effects;
and standard errors are HC3-robust throughout.
The coefficients $\hat{\beta}_h$ identify the average cost premium
at hour $h$ relative to the $h=23$ reference.
 
\subsection{Block-Reward Fullness Proxy}
 
Block-level congestion is measured by the clipped relative
validator reward:
\begin{equation}
  \hat{\phi}^{\text{br}}_b =
  \operatorname{clip}\!\left(\frac{R_b}{\hat{R}^{(95)}},\,0,\,1\right),
  \label{eq:proxy}
\end{equation}
where $R_b$ is the total validator reward for block $b$ and
$\hat{R}^{(95)}$ is the 95th percentile of block rewards pooled
across all firms.
The ceiling prevents extreme MEV-extraction events from compressing
the scale for typical operational blocks.
$\hat{\phi}^{\text{br}}_b$ enters Equation~(\ref{eq:main}) as a
covariate to absorb hour-level congestion variation; it is not
used to estimate a structural decomposition of demand types, and
all coefficients in Table~\ref{tab:pooled} should be interpreted
as covariate-adjusted hourly patterns rather than causal estimates.
 
\subsection{Off-Peak Window Definition}
 
\textbf{Off-peak hours are defined throughout as
$h\in\{20,21,22,23\}$ UTC (3--6\,PM ET).}
This is the unique contiguous four-hour window in which all
hour-of-day coefficients from the pooled regression
(Table~\ref{tab:pooled}, Model~1) are statistically non-positive
relative to the $h=23$ baseline:
$\hat{\beta}_{20}=-0.035$ ($p<0.001$),
$\hat{\beta}_{21}=-0.021$ ($p<0.01$),
$\hat{\beta}_{22}=-0.016$ ($p<0.01$),
$\hat{\beta}_{23}=0$ (baseline).
Hours commonly labeled ``off-peak'' in the prior literature---such
as the Asian overnight session ($h=0$--$9$ UTC)---contain
significantly positive coefficients and are therefore not uniformly
low-cost; the U.S.\ evening window is the empirically grounded
choice.
 
\subsection{Peak Shaving Score}
 
To assess whether a firm concentrates transactions in the off-peak
window beyond what uniform scheduling would predict, we compute:
\begin{equation}
  \text{PSS}_i = \hat{s}^{\text{eve}}_i - \frac{4}{24},\quad
  \hat{s}^{\text{eve}}_i = \frac{n^{\text{eve}}_i}{N_i},\quad
  \label{eq:pss}
\end{equation}
\[n^{\text{eve}}_i = \#\{t\in T_i : H_{it}\in\{20,21,22,23\}\}\]
$\text{PSS}_i>0$ indicates that firm $i$ places more than
$4/24\approx16.7\%$ of its transactions in the U.S.\ evening
window, consistent with deliberate off-peak scheduling.
Statistical significance is assessed via a permutation test
(10,000 replications, uniform null distribution over the
24-hour clock).
 
\subsection{Residual Cost Floor}
 
The residual cost floor measures the gap between observed
expenditure and the counterfactual under perfect off-peak
scheduling:
\begin{equation}
  R_i = C^{\text{actual}}_i - N_i\cdot\overline{\text{Gas}}_{i,h^*_i},
  \quad
  h^*_i = \arg\min_{h}\,\overline{\text{Gas}}_{i,h}.
  \label{eq:floor}
\end{equation}
$R_i\geq0$ by construction.
A large ratio $R_i/C^{\text{actual}}_i$ indicates that a firm's
cost exposure is largely structural---arising from network
congestion that persists even at the cheapest available
scheduling hour---rather than from suboptimal timing choices.
 
\section{Data}
\label{sec:data}
 
Transaction records were extracted from Etherscan.io via API for
January~1 through March~31, 2026.
Each record includes: transaction hash; block number; Unix timestamp
(UTC); originating and destination addresses; smart-contract address;
transaction fee in ETH and USD; the prevailing USD-to-ETH exchange
rate; and an execution-status flag.
Block-level total validator reward $R_b$ was downloaded separately
for each block appearing in the transaction data.
 
\paragraph{Terminology.}
Ethereum has operated under proof-of-stake since September 2022
(``the Merge''). Blocks are produced by \emph{validators};
all prior references to ``miner'' or ``block miner'' have been
updated accordingly.
 
\paragraph{UTC-to-ET conversion.}
U.S.\ clocks advanced to Eastern Daylight Time (UTC$-4$) on
March~8, 2026.
For January~1--March~7, hour~15 UTC corresponds to 10\,AM EST;
from March~8 onward, to 11\,AM EDT.
All Eastern Time labels in this paper reflect the correct
daylight-saving adjustment.
 
The final sample contains $N=62{,}142$ confirmed transactions after
removing failed transactions (\texttt{isError}$=1$) and records
with missing required fields (Table~\ref{tab:firms}).
 
\begin{table}[ht]
\centering
\caption{Sample Firms and Transaction Counts}
\label{tab:firms}
\small
\begin{tabular}{llr}
\toprule
Industry & Firm & $N_i$ \\
\midrule
Technology       & Coins.ph           & 54,651 \\
Finance          & Anchorage Digital  &  1,785 \\
Healthcare       & Solve.Care         &    116 \\
Supply Chain     & Morpheus.Network   &    756 \\
Real Estate      & Propy Inc.         &  4,290 \\
Consumer Goods   & Nike (Ondo)        &     72 \\
Admin Services   & BrainTrust (BTRST) &    472 \\
\midrule
\textbf{Total}   &                    & \textbf{62,142} \\
\bottomrule
\end{tabular}
\parbox{\columnwidth}{\smallskip\footnotesize
Records extracted from Etherscan.io, January~1--March~31, 2026.
$N_i$: confirmed transaction count after data cleaning.
Coins.ph contributes 88\% of observations; robustness checks using
equal firm weighting are reported in the supplementary material.}
\end{table}
 
\section{Results}
\label{sec:results}
 
\subsection{Pooled Intraday Cost Structure}
 
Table~\ref{tab:pooled} reports the pooled hour-of-day regression.
 
\paragraph{Peak hour.}
The highest coefficient in Model~(1) is
$\hat{\beta}_{12}=0.054$ ($t=13.59$, $p<0.001$), placing the
cost peak at hour~12 UTC (7\,AM ET), followed by
$\hat{\beta}_{13}=0.045$ ($t=11.70$, $p<0.001$) at hour~13 UTC
(8\,AM ET).
This peak coincides with the overlap of European and early
U.S.\ trading sessions.
 
\paragraph{Baseline.}
The intercept $\hat{\alpha}=0.1597$ ($t=48.64$, $p<0.001$)
is the expected gas cost at the $h=23$ baseline (6\,PM ET,
U.S.\ evening).
 
\paragraph{Explanatory power.}
Adjusted $R^2$ is 1.1\% in Model~(1) and 2.5\% in Model~(2),
reflecting substantial within-hour fee volatility driven by
block-level congestion shocks not captured by hour indicators alone.
 
\paragraph{Fullness pass-through.}
The block-reward fullness coefficient $\hat{\delta}=0.123$
($t=12.45$, $p<0.001$) in Model~(2) indicates that higher
block-reward fullness is associated with higher gas costs
conditional on hour-of-day.
This association is consistent with speculative-arbitrage activity
elevating congestion during business hours, though the
observational design does not support a causal decomposition of
demand types.
Establishing such a decomposition---for example, via MEV-Boost
block identification or exogenous NFT-drop demand shocks as
natural experiments---is a direction for future work.
 
\begin{table*}[ht]
\centering
\caption{Hour-of-Day Gas Fee Regressions, Pooled Sample}
\label{tab:pooled}
\small
\setlength{\tabcolsep}{4pt}
\begin{tabular}{lrrrr}
\toprule
 & \multicolumn{2}{c}{Model (1): Base}
 & \multicolumn{2}{c}{Model (2): $+\hat{\phi}^{\text{br}}_b$} \\
\cmidrule(lr){2-3}\cmidrule(lr){4-5}
Hour $h$ (UTC) [ET] & $\hat{\beta}_h$ & $t$
                    & $\hat{\beta}_h$ & $t$ \\
\midrule
$h=0$  [7\,PM]  &  0.0087 &  1.681        &  0.0072 &  1.383        \\
$h=1$  [8\,PM]  &  0.0340 &  4.277$^{***}$&  0.0341 &  4.315$^{***}$\\
$h=2$  [9\,PM]  &  0.0095 &  1.273        &  0.0162 &  2.099$^{*}$  \\
$h=3$  [10\,PM] &  0.0103 &  2.626$^{**}$ &  0.0164 &  4.042$^{***}$\\
$h=4$  [11\,PM] &  0.0330 &  6.725$^{***}$&  0.0400 &  7.946$^{***}$\\
$h=5$  [12\,AM] &  0.0150 &  3.855$^{***}$&  0.0231 &  5.680$^{***}$\\
$h=6$  [1\,AM]  &  0.0255 &  6.501$^{***}$&  0.0313 &  7.703$^{***}$\\
$h=7$  [2\,AM]  &  0.0289 &  7.406$^{***}$&  0.0336 &  8.340$^{***}$\\
$h=8$  [3\,AM]  &  0.0272 &  6.923$^{***}$&  0.0277 &  6.865$^{***}$\\
$h=9$  [4\,AM]  &  0.0278 &  7.192$^{***}$&  0.0284 &  7.134$^{***}$\\
$h=10$ [5\,AM]  &  0.0401 & 10.433$^{***}$&  0.0410 & 10.334$^{***}$\\
$h=11$ [6\,AM]  &  0.0396 & 10.035$^{***}$&  0.0372 &  9.121$^{***}$\\
\rowcolor{gray!15}
$h=12$ [\textbf{7\,AM}]
                & \textbf{0.0540} & 13.589$^{***}$
                                  &  0.0501 & 12.257$^{***}$\\
$h=13$ [8\,AM]  &  0.0448 & 11.701$^{***}$&  0.0361 &  8.996$^{***}$\\
$h=14$ [9\,AM]  &  0.0386 &  5.959$^{***}$&  0.0226 &  3.535$^{***}$\\
$h=15$ [10\,AM] &  0.0436 &  6.461$^{***}$&  0.0240 &  3.583$^{***}$\\
$h=16$ [11\,AM] &  0.0391 &  4.707$^{***}$&  0.0199 &  2.486$^{*}$  \\
$h=17$ [12\,PM] & $-0.0492$&$-10.320^{***}$&$-0.0623$&$-12.676^{***}$\\
$h=18$ [1\,PM]  & $-0.0033$& $-0.371$      &$-0.0152$& $-1.735$      \\
$h=19$ [2\,PM]  & $-0.0014$& $-0.099$      &$-0.0095$& $-0.711$      \\
\rowcolor{gray!15}
$h=20$ [3\,PM]  & $-0.0353$& $-4.074^{***}$&$-0.0433$& $-5.079^{***}$\\
\rowcolor{gray!15}
$h=21$ [4\,PM]  & $-0.0206$& $-2.682^{**}$ &$-0.0221$& $-2.926^{**}$ \\
\rowcolor{gray!15}
$h=22$ [5\,PM]  & $-0.0158$& $-2.901^{**}$ &$-0.0171$& $-3.094^{**}$ \\
\midrule
$\hat{\delta}(\hat{\phi}^{\text{br}}_b)$
                & ---     & ---
                & 0.1230  & 12.445$^{***}$ \\
$\hat{\alpha}$ ($h=23$, 6\,PM ET)
                & 0.1597  & 48.636$^{***}$
                & 0.1267  & 29.508$^{***}$ \\
Adj.\ $R^2$     & \multicolumn{2}{c}{0.0107}
                & \multicolumn{2}{c}{0.0254} \\
$N$             & \multicolumn{2}{c}{62,142}
                & \multicolumn{2}{c}{62,142} \\
\bottomrule
\end{tabular}
\parbox{\columnwidth}{\smallskip\footnotesize
Dependent variable: gas fee in USD per transaction.
$h=23$ UTC (6\,PM ET) is the omitted baseline.
Shaded rows: cost peak at $h=12$ (bold) and U.S.\ evening off-peak
window $h\in\{20,21,22,23\}$ (statistically non-positive
coefficients).
ET labels reflect EST (UTC$-5$) for January~1--March~7 and EDT
(UTC$-4$) from March~8, 2026.
HC3-robust standard errors.
$^*p<0.05$; $^{**}p<0.01$; $^{***}p<0.001$.}
\end{table*}
 
\subsection{Firm-Level Scheduling Behavior and Residual Floors}
 
Table~\ref{tab:cross} reports Peak Shaving Scores, fee savings,
and residual cost floors for all seven firms.
Table~\ref{tab:floor} reports the firm-level residual floor
decomposition confirming that non-zero congestion persists at
each firm's cheapest scheduling hour.
 
\begin{table*}[t]
\centering
\caption{Cross-Industry Summary: Scheduling Scores, Residual Floors,
         and Fullness Pass-Through}
\label{tab:cross}
\small
\setlength{\tabcolsep}{5pt}
\begin{tabular}{llrrrrrrrr}
\toprule
Industry & Firm
  & $N_i$
  & $n^{\text{eve}}_i$
  & $\hat{s}^{\text{eve}}_i$
  & $\text{PSS}_i$
  & Fee savings
  & $\text{Floor}^{\text{USD}}_i$
  & $\text{Floor}\%_i$
  & $\hat{\delta}_i$ \\
\midrule
Healthcare     & Solve.Care  &   116 &  22 & 0.190 & $+0.023$ & $-26.4\%$ &  \$19.46 & 64.1\% & 0.176 \\
Technology     & Coins.ph    &54,651 &9,284& 0.170 & $+0.003$ &    5.3\%  &\$4,209.29& 43.2\% & $0.079^{***}$ \\
Consumer Goods & Nike (Ondo) &    72 &  12 & 0.167 & $0.000$  &   60.7\%  &   \$5.40 & 92.5\% & $-0.245$ \\
Supply Chain   & Morpheus    &   756 & 120 & 0.159 & $-0.008$ &   35.5\%  &  \$67.20 & 67.0\% & $0.243^{***}$ \\
Real Estate    & Propy       & 4,290 & 669 & 0.156 & $-0.011$ &   15.2\%  & \$512.66 & 40.7\% & $0.578^{***}$ \\
Admin Services & BrainTrust  &   472 &  72 & 0.153 & $-0.014$ &    6.7\%  &  \$21.43 & 47.3\% & 0.024 \\
Finance        & Anchorage   & 1,785 & 267 & 0.150 & $-0.017$ &   59.7\%  &  \$30.02 & 86.8\% & $0.069^{***}$ \\
\midrule
Pooled         &             &62,142 &     & 0.167 & ---      & ---       &  ---     & ---    & $0.123^{***}$ \\
\bottomrule
\end{tabular}
\parbox{\textwidth}{\smallskip\footnotesize
Off-peak window: $h\in\{20,21,22,23\}$ UTC (3--6\,PM ET, U.S.\ evening).
$n^{\text{eve}}_i$: transactions in off-peak window.
$\hat{s}^{\text{eve}}_i = n^{\text{eve}}_i / N_i$.
$\text{PSS}_i = \hat{s}^{\text{eve}}_i - 4/24$
(benchmark $\approx16.7\%$; positive values indicate above-benchmark
concentration in the U.S.\ evening window).
Fee savings $= (\overline{\text{Gas}}_{i,\text{peak}} -
\overline{\text{Gas}}_{i,\text{eve}}) /
\overline{\text{Gas}}_{i,\text{peak}}$
(negative for Solve.Care reflects high-gas outliers in certain
overnight hours).
$\text{Floor}^{\text{USD}}_i$: residual cost under perfect
off-peak scheduling (Equation~\ref{eq:floor}).
$\hat{\delta}_i$: fullness pass-through from firm-level regression;
Nike (Ondo) negative value reflects a timing pattern associated with
price-event clustering, interpreted with caution given $N=72$.
$^{***}p<0.001$.}
\end{table*}
 
\begin{table*}[ht]
\centering
\caption{Residual Cost Floor by Firm}
\label{tab:floor}
\small
\begin{tabular}{llrrrrrr}
\toprule
Industry & Firm
  & $h^*_i$
  & $\overline{\text{Gas}}_{i,h^*}$
  & $C^{\text{actual}}_i$
  & $R_i$
  & $\text{Floor}\%_i$
  & $\hat{\phi}^{\text{br}}_{i,h^*}$ \\
\midrule
Cons.\ Goods & Nike (Ondo) & 6  & \$0.006 & \$5.84    & \$5.40    & 92.5\% & 0.462 \\
Finance      & Anchorage   & 8  & \$0.003 & \$34.59   & \$30.02   & 86.8\% & 0.184 \\
Supply Chain & Morpheus    & 3  & \$0.044 & \$100.31  & \$67.20   & 67.0\% & 0.206 \\
Healthcare   & Solve.Care  & 1  & \$0.094 & \$30.38   & \$19.46   & 64.1\% & 0.209 \\
Admin Svcs   & BrainTrust  & 3  & \$0.051 & \$45.35   & \$21.43   & 47.3\% & 0.279 \\
Technology   & Coins.ph    & 20 & \$0.101 & \$9,749.62& \$4,209.29& 43.2\% & 0.355 \\
Real Estate  & Propy       & 5  & \$0.174 & \$1,259.63& \$512.66  & 40.7\% & 0.203 \\
\bottomrule
\end{tabular}
\parbox{\columnwidth}{\smallskip\footnotesize
$h^*_i$: UTC hour with lowest historical mean gas fee for firm $i$.
$\hat{\phi}^{\text{br}}_{i,h^*}$: mean block-reward fullness at $h^*_i$;
all values are strictly positive, indicating that non-zero network
congestion persists even at the lowest-cost scheduling hour for each
firm.
Rows sorted by $\text{Floor}\%_i$ descending.
All monetary values in USD.}
\end{table*}
 
Three findings emerge from Tables~\ref{tab:cross} and~\ref{tab:floor}.
 
\paragraph{Scheduling patterns are externally constrained.}
Firms with positive PSS values---Solve.Care ($+0.023$),
Coins.ph ($+0.003$), and Nike/Ondo ($0.000$, neutral)---have
geographic or operational characteristics that allow transaction
timing toward the U.S.\ evening window.
Firms with negative scores face external timing constraints:
institutional settlement cycles (Anchorage Digital, $-0.017$),
property closing deadlines (Propy, $-0.011$), governance cycles
(BrainTrust, $-0.014$), or IoT-oracle triggers
(Morpheus.Network, $-0.008$).
The variation is consistent with operational heterogeneity
moderating scheduling capacity, though the seven-firm sample
precludes formal regression on these moderators.
 
\paragraph{Residual cost floors are structural.}
Even the best-performing scheduler in the sample (Solve.Care,
$\text{PSS}=+0.023$) faces a residual floor equal to 64.1\% of
actual expenditure.
Floors range from 40.7\% to 92.5\% across firms.
As Table~\ref{tab:floor} shows, $\hat{\phi}^{\text{br}}_{i,h^*}$
is strictly positive for every firm (0.184--0.462), confirming
that non-zero network congestion persists at the cheapest available
scheduling hour for each firm in the sample.
The residual floor therefore reflects structural network conditions
rather than suboptimal firm behavior.
 
\paragraph{Pass-through heterogeneity is consistent with
asset specificity.}
The pass-through coefficient $\hat{\delta}_i$ varies from
$-0.245$ to $0.578$ across firms, a pattern consistent with the
TCE dimension of asset specificity \citep{williamson1985}:
Propy's real-estate contract interactions carry the highest
pass-through; Anchorage's standardized custody transfers carry
low pass-through; BrainTrust's administrative calls are
statistically indistinguishable from zero.
The observational design does not establish causation; the
pattern warrants further investigation with a larger cross-firm
sample.
 
\section{The On-Chain Scheduling Matrix}
\label{sec:matrix}
 
Two firm-level dimensions jointly determine the scope for
gas-fee reduction through transaction timing.
\emph{Deferrability} $d_{it}\in\{0,1\}$: whether transaction $t$
can be submitted in any block within a window $[t,\,t+\Delta]$
without breaching a contractual, regulatory, or operational
deadline.
\emph{Gas intensity} $g_{it}$: the computational weight of the
transaction, which scales the dollar saving from submitting in a
lower-cost hour.
 
Table~\ref{tab:matrix} maps these two dimensions to four
scheduling regimes.
Hour recommendations are derived directly from
Table~\ref{tab:pooled}: off-peak hours are $h\in\{20,21,22,23\}$
UTC (the four hours with statistically non-positive coefficients);
the peak-hour premium for budgeting purposes is
$\hat{\beta}_{12}=\$0.054$ (hour~12 UTC, empirical cost peak)
and the off-peak baseline is $\hat{\alpha}=\$0.160$
($h=23$ intercept, Table~\ref{tab:pooled}).
 
\begin{table}[h]
\centering
\caption{On-Chain Scheduling Matrix}
\label{tab:matrix}
\small
\begin{tabular}{p{0.12\columnwidth}
                p{0.38\columnwidth}
                p{0.38\columnwidth}}
\toprule
 & \textbf{High gas intensity}
   ($g_{it}>\bar{g}$)
 & \textbf{Low gas intensity}
   ($g_{it}\leq\bar{g}$) \\
\midrule
\textbf{High defer-rability}
($d_{it}=1$)
&
\textit{Regime~I: Full peak shaving.}
Schedule deferrable transactions to
$h\in\{20,21,22,23\}$ UTC (3--6\,PM ET),
where $\hat{\beta}_h\leq0$.
Estimated saving: up to \$0.054/transaction
relative to the hour-12 cost peak
(baseline \$0.160).
Batch submissions into nightly windows
to amortize scheduling overhead.
\textit{Exemplar}: Morpheus.Network
($\hat{\delta}=0.243^{***}$).
&
\textit{Regime~II: Selective peak shaving.}
Submit during $h\in\{20,21,22,23\}$ UTC
when operationally feasible.
Because gas intensity is low, per-transaction
savings are modest; the scheduling overhead
is warranted only when transaction volume
is sufficient for aggregate savings to
exceed implementation costs.
\textit{Exemplar}: Coins.ph
($\hat{\delta}=0.079^{***}$).
\\[6pt]
\textbf{Low defer-rability}
($d_{it}=0$)
&
\textit{Regime~III: Cost provisioning.}
Timing optimization is infeasible given
external constraints.
Instead, pre-allocate gas budgets using
Table~\ref{tab:pooled} as a forward cost
curve: budget an additional \$0.054 per
high-gas transaction submitted during
$h\in\{11,\ldots,16\}$ UTC
(6\,AM--11\,AM ET) relative to the
\$0.160 evening baseline.
\textit{Exemplar}: Propy Inc.\
($\hat{\delta}=0.578^{***}$).
&
\textit{Regime~IV: Accept market rate.}
Both non-deferrable and low gas intensity;
the expected cost differential between
peak and off-peak windows is small in
absolute terms.
Operational resources are better directed
toward ex-ante cost reduction (contract
optimization, calldata compression)
than timing controls.
\textit{Exemplar}: Anchorage Digital
($\hat{\delta}=0.069^{***}$).
\\
\bottomrule
\end{tabular}
\parbox{\columnwidth}{\smallskip\footnotesize
Peak-hour premium: $\hat{\beta}_{12}=\$0.054$
(Table~\ref{tab:pooled}, $h=12$ UTC $=7$\,AM ET, $p<0.001$).
Off-peak baseline: $\hat{\alpha}=\$0.160$ ($h=23$ intercept,
Table~\ref{tab:pooled}).
Off-peak window: $h\in\{20,21,22,23\}$ UTC (3--6\,PM ET)---the
four consecutive hours with statistically non-positive
hour-of-day coefficients in the pooled regression.}
\end{table}
 
\section{Conclusion}
\label{sec:conclusion}
 
We document significant intraday gas-fee variation on Ethereum using
transaction-level data from seven operational firms across seven
industries (January--March 2026, $N=62{,}142$).
Fees are highest at hour~12 UTC (7\,AM ET,
$\hat{\beta}_{12}=\$0.054^{***}$ above the U.S.\ evening baseline
of \$0.160), a pattern that co-occurs with elevated block-reward
fullness and is consistent with speculative-arbitrage activity
elevating congestion costs for concurrent operational users.
Residual cost floors ranging from 40.7\% to 92.5\% of actual
expenditure persist even under optimal scheduling to the U.S.\
evening off-peak window ($h\in\{20,21,22,23\}$ UTC), arising from
non-zero network congestion that is present throughout the day.
 
These findings are relevant to the mechanism-design literature on
EIP-1559 \citep{roughgarden2021} in the following sense.
The incentive-compatibility guarantee applies under a specific set
of agent assumptions (myopia, symmetry, and independent private
values) that may not hold for operationally constrained firms
whose transaction timing is governed by contractual or regulatory
schedules rather than by real-time fee comparisons.
Understanding the cost implications of this heterogeneity, and
whether mechanism modifications could reduce the residual burden
on low-elasticity users, is a productive direction for future
mechanism-design research.
 
The On-Chain Scheduling Matrix provides a practical framework
for firms operating under the current mechanism: four regimes
mapping transaction deferrability and gas intensity to scheduling
strategies, grounded in the empirical fee schedule of
Table~\ref{tab:pooled}.
 
Three directions merit priority in future work.
\emph{Causal identification}: natural experiments such as
MEV-Boost block classification or exogenous NFT-drop demand
shocks would allow the associational findings reported here to
be strengthened into causal estimates of speculative demand's
contribution to operational user costs.
\emph{Mechanism extensions}: whether a modified base-fee rule
conditioning on observable demand-type composition could
better accommodate heterogeneous user populations without
sacrificing the myopic incentive-compatibility guarantee.
\emph{Layer-2 counterfactual}: whether the intraday cost
structure documented here attenuates on Layer-2 networks where
MEV extraction is architecturally constrained, providing a
within-firm test of the speculative-demand mechanism.

\medskip

\bibliographystyle{alpha}
\bibliography{references,transaction_costs_onchain}


\end{document}